# DIVERGENCE OF TIMEOUT ALGORITHMS FOR PACKET RETRANSMISSIONS

Raj Jain

Digital Equipment Corp.
77 Reed Road (HLO2-3/N03)
Hudson, MA 01749

## ABSTRACT

The problem of adaptively setting the timeout interval for retransmitting a packet has been discussed. A layered view of the algorithms has been presented. It is shown that a timeout algorithm consists of essentially five layers or procedures which can be independently chosen and modified. A number of timeout algorithms proposed in the literature have been decomposed into these five layers.

One of the key layers not discussed in the literature is that of determining the sample round trip delay for packets that have been transmitted more than once. It is shown that this layer has a significant impact on the network performance.

Under repeated packet loss, most timeout algorithms either diverge or converge to a wrong value. A number of alternative schemes have been presented. It is argued that divergence is preferable to false convergence. It is a feature that is helpful in reducing network traffic during congestion.

## Overview

Almost all networking protocols have to cope with the problem of determining when a node should retransmit a packet that has not been acknowledged. The timeout algorithm plays an important role in the overall stability and performance of networking protocols. A bad timeout algorithm, e.g., quick timeouts, may flood a network with duplicate copies of packets and lead to unwanted congestion. In fact, it is during congestion that a good timeout algorithm really pays off.

Determining an optimal timeout interval is important, not only for computer networks but also for other distributed systems [12,13] and applications, e.g., distributed databases, remote procedure calls, process-to-process communication, file servers, print servers. In these cases, the problem is only slightly different. Here we concentrate mainly on computer networks, but the ideas can be adapted easily for these other applications. We assume that the link errors are handled by a suitable link level protocol and timeouts are used to detect packet losses at the end-to-end (transport) or higher layers.

A number of timeout algorithms have been proposed and analyzed in literature [1,4,5,11,14-17]. In this paper we present a layered view of timeout algorithms. A number of alternative procedures have been presented for each layer. By choosing an appropriate procedure for each layer, an architect can design a timeout algorithm that is suitable for a particular network.

One problem that has not been addressed in the literature is the divergence of the timeout algorithms as a result of their recursive nature. The measured round-trip delays usually include the timeout intervals used for retransmissions, and the timeout intervals are calculated based on measured round-trip delays. Thus, larger timeout intervals lead to larger round-trip delay estimates, and repeated retransmissions result in a positive feedback system that diverges. This may reduce the throughput to zero.

This paper discusses the problem of divergence and shows that most timeout algorithms either diverge or converge to a wrong value. The latter phenomenon, called false convergence is dangerous because it may lead to unnecessary congestion. We conclude that between the two choices, divergence is preferable in that it helps avoid congestion.

With the intention of conveying our message to network implementors and designers, we have kept the mathematical sophistication to a minimum and tried to explain our results with illustrations.

## Layered View of a Timeout Algorithm

The simplest solution to the timeout problem is to have a static algorithm with a fixed timeout interval. Some of the parameters required for determining the optimal timeout interval are listed in table 1.

Table 1: List of network characteristics that impact the timeout interval

| | |
|---|---|
| 1. | Distribution of packet sizes. |
| 2. | Link/nodes speeds on the path. |
| 3. | Service time distribution at nodes: source, destination, gateways. |
| 4. | Other activities at nodes. |
| 5. | Other traffic sharing the path. |
| 6. | Number of buffers at nodes. |
| 7. | Flow control window sizes. Larger window sizes cause longer queueing delay and require longer timeout intervals. |

Although given parameters from table 1, one can calculate optimal timeout interval [1,5,11,15,16], a problem exists in that none of the parameters is generally known and many of these vary dynamically with time. The dynamic nature of parameters makes an adaptive timeout algorithm a virtual necessity.

The key functions of an adaptive timeout algorithm are to:

1. Estimate current round-trip delay
2. Detect transient failures (packet loss) which can be recovered by simply retransmitting the packet.
3. Help reduce input to the network if congestion (repeated packet loss) is sensed.
4. Detect permanent failures (broken paths or nodes) which require circuit disconnection.

An adaptive timeout algorithm should therefore have at least four distinct procedures or components corresponding to the above functions. The function of round-trip delay estimation consists of two different functions:

a. Estimate round-trip delay based on packets which are acknowledged without any retransmission.

b. Estimate round-trip delay for packets which require one or more retransmissions.

The procedures for these two functions may not necessarily be the same. Thus, adaptive timeout algorithms consist of five components in all. Each of these components acts as a layer, in the sense that, upper layers use the services of the lower layers, lower layers are independent of the upper layers. There are many alternatives for each layer which can be independently chosen without modifying upper layers. This layered view is shown in Figure 1. In this section, we describe the function of each layer. In the next section, we describe several alternatives for each layer. We use the term procedures for alternatives for these layers. Thus, selecting one procedure for each of the five components results in a timeout algorithm.

The five layers are as follows:

| 5. Disconnection |
| --- |
| 4. Back off |
| 3. First Timeout Computation |
| 2. Delay Estimation with retransmission |
| 1. Delay Estimation w/o Retransmission |

Figure 1: Adaptive timeout algorithms consist of five independent layers.

1. Round Trip Delay Estimation Without Retransmissions: A procedure $F_1$ is used to update the round-trip delay estimate $E$ based on measured sample delay $S$ for packets that are acknowledged without any retransmissions.

$$E \leftarrow F_1(E,S)$$

2. Round Trip Delay Estimation with Retransmissions: If a packet is transmitted more than once, it may be necessary to use a different procedure to update the round-trip delay estimate.

$$E \leftarrow F_2(E,S)$$

As we will see later, this is the most interesting layer as it has a significant impact on the stability of the network. This is because measuring the sample delay $S$ is not straight-forward in this case.

3. First Timeout Interval Computation: Based on the current round-trip delay estimate, the sender times-out the first time usually using the following formula:

$$t_0 \leftarrow F_3(E)$$

4. Back-off: If a packet has been retransmitted once and the copy gets lost, the second and subsequent timeout interval may be calculated using a back-off procedure, such as the following:

$$t_i \leftarrow F_4(E,t_0,t_1,...,t_{i-1}) \quad i=1,2,3,...$$

Here, $t_i$ is the timeout interval after $i^{th}$ retransmission.

5. Disconnection: This procedure helps decide if a permanent failure has occurred and the circuit needs to be disconnected.

IF $F_5(E,t_0,t_1,...,t_i,i) = True$ THEN Disconnect

In the next five sections, we discuss alternatives for each of these five layers in more detail.

Layer 1: Round Trip Delay Estimation Without Retransmissions

A commonly used procedure is the exponentially weighted average:

$$E \leftarrow \alpha E + (1-\alpha)S$$

Here $E$ is the current estimate, $S$ is the sample delay measured between transmission of a packet and reception of its ack, and $\alpha$ is a parameter which can be set between 0 and 1. If $\alpha$ is set at $1-2^{-n}$ for some $n$, the update reduces to a simple subtract and shift as shown below:

$$E \leftarrow E + 2^{-n}(S-E)$$

The parameter $\alpha$ determines the weighting placed on the old estimate. A large value for $\alpha$ implies that the old value will be weighted more heavily and the recent variations in the network delay will have little impact on the estimate. The estimate does not respond very quickly to short term variations. It takes quite a few samples in the same direction (up or down) before the estimate picks that value. Thus, the estimate is more stable.

A very small value for $\alpha$ implies heavier weight for the most recent sample. The estimate responds quickly to changes in network conditions. Sometimes, however, this response is quicker that it ought to be. For example, consider a lucky packet which reached the destination very quickly. A low $\alpha$ will cause the estimate to come down immediately, based on one sample. The sender uses this low estimate to timeout the next packet even though the packet is well on its way to the destination.

Mills [14] proposes using two values $\alpha_1$ and $\alpha_2$ depending upon whether the sample is less or greater than the current estimate:

IF $S<E$ THEN $E \leftarrow \alpha_1 E + (1-\alpha_1)S$
ELSE $E \leftarrow \alpha_2 E + (1-\alpha_2)S$

Here, $0<\alpha_2<\alpha_1<1$. The effect is to make the algorithm more responsive to upward trends in delay and less responsive to downward trends. The suggested values of the parameters are $\alpha_1=15/16$, and $\alpha_2=3/4$.

Edge [4] proposes an algorithm which estimates the average $E$ as well as the variance $V$ using exponentially weighted averaging:

$$E \leftarrow \alpha E + (1-\alpha)S$$

$$V \leftarrow \beta V + (1-\beta)(S-E)^2$$

Here, $\beta$ is another parameter between 0 and 1.

Layer 2: Round Trip Delay Estimation with Retransmissions

If a packet is retransmitted, it is not obvious how one can determine the exact round-trip delay for that packet. One way would be to measure the sample delay as the interval between first transmission of the packet and the final receipt of the ack. This sample delay can then be used in the same exponential weighting average formula as above. This leads to a diverging situation in which the estimate keeps increasing. We will discuss this problem in detail later in this paper.

Layer 3: First Timeout Interval Calculation

One commonly used procedure for calculating the first timeout interval is:

$$t_0 \leftarrow kE$$

Where, $E$ is the current estimate of the round-trip delay and the parameter $k$ is chosen to get an acceptable probability of false alarms.

A low value of $k$ results in low timeout intervals which in turn causes too many false retransmissions and duplicate packets leading to sometimes to unnecessary congestion of the network. A very large value of $k$, on the other hand, results in a long timeout interval and hence loss of time and throughput.

The probability of false alarms is essentially the probability of sample round-trip delay being more than $k$ times the estimate. Hence the optimal value of $k$ is a function of variance of the delay, cost of duplicate packets, and the cost of lost time.

It is suggested that low values of $k$ be used for lightly loaded networks because the cost of duplicate packets is rather low on such networks. A high value for $k$ should be used for congested networks because in this case the duplicate packets may cause more congestion in addition the time lost by one source can be used by other sources on the network.

Edge's algorithm [4] uses both the estimated mean and the variance:

$$t_0 \leftarrow E + k\sqrt{V}$$

### Layer 4: Back-off procedure

If a packet needs to be retransmitted more than once, this may indicate severe congestion in the network. Increased timeout interval in such a case may help reduce the congestion. Some of the possibilities for calculating timeout interval after the first retransmission are:

1. No back-off: All retransmission use the same timeout interval.

$$t_i \leftarrow t_0$$

2. Exponential back-off: The timeout intervals are increased exponentially.

$$t_i \leftarrow b t_{i-1}$$

The parameter $b$ controls the rate of increase.

3. Random exponential back-off: The timeout intervals are randomly generated with range increasing exponentially.

$$t_i \leftarrow Uniform(t_{min}, b^i k t_0)$$

This is similar to the binary exponential back-off scheme used in Ethernet™ for retransmissions after a collision.

3. Linear back-off: The timeout interval is increased linearly.

$$t_i \leftarrow t_{i-1} + \Delta t$$

Any of the above procedures may also be used with a threshold, i.e., the timeout intervals are not increased beyond a certain maximum $t_{max}$.

$$t_i \leftarrow min \{ t_{max}, Calculated\ Interval \}$$

The purpose of the threshold is to prevent the timeout algorithm from diverging into an unrealistic range. However, our simulations show that the thresholds do not help, because if the algorithm diverges the situation must be really bad and retrying at threshold is not going to make it any better.

### Layer 5: Disconnection Procedure

The purpose of this procedure is to declare a dead or almost dead (highly congested) node as dead. The aim is to minimize the probability of false disconnect and also to minimize the time to discover a dead node. However, the goals are different at various stages of connection. For example, at connection set up, declaring a working node as down does not hurt as much as it does in the middle of a connection after several thousand packets have already been received. At connection set-up, the procedure can be pessimistic, i.e., without too many tries, it can declare a path as down and the loss that is due to a false declaration is very small. After some packets have been transmitted, the disconnection procedure should be optimistic and should not err on the side of declaring a correctly operating link down. It should try its best to make sure that the link is in fact broken, even if that means many retries.

The two cases are discussed separately below.

1. Disconnection Procedure At Setup Time: As shown in Figure 2, one

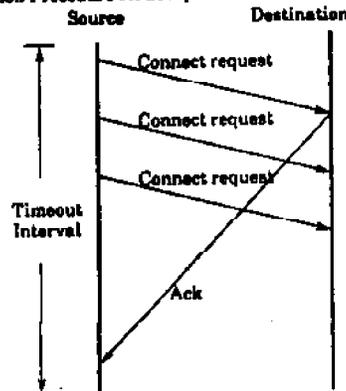

Figure 2: At connection setup time, one may quickly send many copies of the connect request and thus minimize the time to discover a broken path.

possibility is to use a small timeout interval $t_0$ and send $r$ copies of the connect request one after another and then wait for an acknowledgment to arrive for any one of them. The purpose of sending the $r$ copies quickly is to minimize the time needed to detect a failed or congested node. If the ack does not arrive after a threshold period, then the procedure should be retried. The threshold is set based on human patience, for example, a few seconds. Also, at this point the user should be given an appropriate indication, such as "retrying ...". This will allow the user to interrupt and cancel or reschedule the request if necessary.

2. Disconnection Procedure During a Connection: Some of the possibilities at this point are:

a. Disconnect after $r$ retries. Here $r$ is a fixed number. Small values of $r$ will result in frequent disconnects, and large values will cause too many unnecessary packets to be injected in the network before discovering that the destination is not reachable. The network performance is not very sensitive to $r$ as long as it is not set too low and nodes do not fail with a high frequency. A good default value is 10.

b. Disconnect after $r$ retries, but increase $r$ with the number of packets that have been successfully sent so far.

c. Disconnect after the total sum of timeout intervals has exceeded a threshold <u>or</u> at least $r$ retries have been made. This sets a limit on the total time lost if a path breaks down.

d. Disconnect after the total sum of timeout intervals has exceeded a threshold <u>and</u> at least $r$ retries have been made. The threshold is set based on the possibility that one of the nodes on the path fails and reboots itself. For example, if the time to reboot the node is 20 seconds, we may want to wait at least 20 seconds before giving up on a lost packet.

### Summary of Proposed Algorithms

A number of timeout algorithms have been proposed and analyzed in the literature. Some [1, 5, 11, 15, 16] discuss static timeout and determine the optimal timeout value. Others [4,14,17] analyze adaptive timeout algorithms. This later set of algorithms have been decomposed into layers and shown in Table 2.

### Divergence of the Timer Algorithm

As seen from Table 2, the second layer of measuring the delay for packets incurring retransmission has not been explicitly discussed except in the case of algorithm A1. This is the algorithm currently in use in DECnet's™ NSP as well as ARPAnet's TCP protocols.

The procedure chosen for this layer has a significant impact on the divergence of the algorithm. All timeout algorithms work perfectly if the packets are lost only occasionally, which is usually the case. However, under sustained congestion conditions and repeated packet losses, the algorithms may diverge, i.e., the timeout interval may increase and throughput may drop to zero.

A diverging situation is illustrated by an example below. Let us consider the algorithm A1 with the following parameter values:

1. Timeout multiplier $k = 4$.
2. Exponential averaging weight $\alpha = 0.5$
3. Actual round-trip delay is constant at one.
4. Initial estimate of the delay $E_0$ is also one.

We assume that the alternate packets are lost, and thus each packet needs to be transmitted twice. The events are explained in Figure 3. From it we see that the delay estimate $E_i$ after $i$ updates is

$$E_i = \{4(2.5)^i - 1\}/3$$

The example discussed above is simplified. However, the divergence occurs for almost all alternatives as we will show later in this paper. The maximum allowable packet loss rate depends upon the parameters used. For example, Algorithm A1 diverges if loss probability is more than $1/(1+k)$.

Table 2: A Layered view of adaptive timeout algorithms proposed in the literature.

| Layer | A1 [3,6] | A2 [14] | A3 [17] | A4 [4] |
|---|---|---|---|---|
| 1. Estimation without Retrans. | $E \leftarrow \alpha E + (1-\alpha)S$ | IF $S < E$ THEN $E \leftarrow \alpha_1 E + (1-\alpha_1)S$ ELSE $E \leftarrow \alpha_2 E + (1-\alpha_2)S$ | $E \leftarrow \alpha E + (1-\alpha)S$ | $V \leftarrow \beta V + (1-\beta)(S-E)^2$ $E \leftarrow \alpha E + (1-\alpha)S$ |
| 2. Estimation with Retrans. | $S \leftarrow$ Time from 1st transmission to 1st ack $E \leftarrow \alpha E + (1-\alpha)S$ | * | * | * |
| 3. 1st Timeout Comput. | $t_0 \leftarrow kE$ | $t_0 \leftarrow kE$ | $t_0 \leftarrow Max(T_{min},$ $min(kE, T_{max}))$ | $t_0 \leftarrow E + k\sqrt{V}$ |
| 4. Back-off | None | None | None | None |
| 5. Disconn. | max $r$ retries | * | * | * |

\* Not explicitly discussed.

### Tsao-Lee Anomaly

The possibility of the network throughput dropping down to zero was first discovered in an experiment conducted by N. Tsao and S. Lee at Digital Equipment Corporation. The experiment showed that increasing the line speeds does not necessarily increase the performance of a network. The experiment, a contrived one, was specially designed to study the effect of too few buffers in the intermediate nodes.

Four nodes serially connected by three 19.2 Kb/s lines were used in the experiment, as shown in Figure 4. All the intermediate nodes were configured with very few buffers. The time to transfer a particular file was measured as five minutes. After the line between the first two nodes was replaced by a fast 1 Mb/s line, the transfer time increased to seven hours!

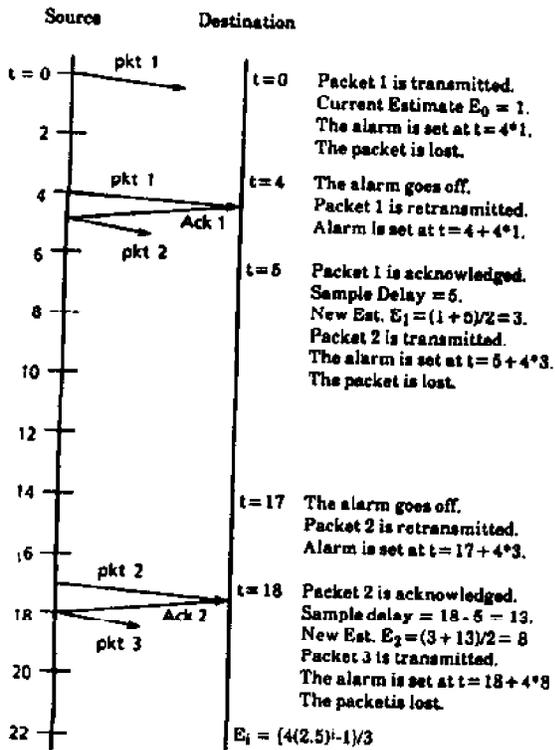

Figure 3: A simple example of divergence of the timeout algorithm A1. In this case, $\alpha = 0.5$, and $k = 4$. It is assumed that alternate packets are lost, the round-trip delay is 1, and the initial estimate $E_0$ is 1. Notice that estimate $E_i$ is an exponentially increasing function.

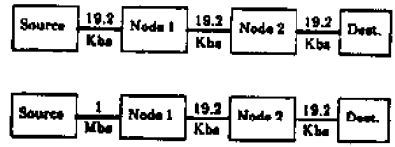

Figure 4: When the 19.2 Kb line from the source was replaced by a 1Mb line, the time to transfer a file increased from five minutes to seven hours due to packet loss at the first intermediate node.

A close examination showed that the CPU utilization was very low, because the source spent most of the time waiting for timeout to occur. In the first configuration, the lines coming in and going out of the first intermediate node were at the same speed. Therefore, by the time a packet arrived, the previous one had left, and the intermediate node never dropped any packet for lack of buffers. In the second configuration, the incoming line to the first intermediate node was so fast that the packets could not be transmitted as quickly as they arrived. Thus the intermediate node soon started dropping packets due to unavailability of buffers. The packet loss caused the source's round-trip delay estimate to increase substantially. This in turn increased the timeout interval, and the source spent most of the time simply waiting for the timeout alarm to go off.

### Alternatives for Delay Estimation During Sustained Loss

In the previous section, we illustrated that the timeout algorithm A1 diverges if the packets are lost repeatedly. One could argue that one way to fix the problem would be to change the round-trip delay estimation algorithm. We tried several algorithms and came to the conclusion that each of these variations had unique problems. In this section, we describe the alternatives that were considered.

The basic problem is that information available at the source is incomplete. Suppose a source receives an acknowledgment for a packet after $n$ transmissions (see Figure 5). If we know that $i^{th}$ copy is being acknowledged, then the correct sample delay is the interval between sending the $i^{th}$ copy and receiving the acknowledgment. Since the acknowledgment does not tell which copy of the lost packet is being acknowledged, it is necessary to arbitrarily assume some value for $i$. In the algorithm as described above, we assumed $i=1$, i.e., the sample delay is always measured from the first attempt. This gives an inflated estimate of the round-trip delay and leads to divergence. There is positive feedback such that an increased delay estimate increases the timeout interval, which in turn increases the delay estimate, and the cycle continues. The solutions that come to mind are:

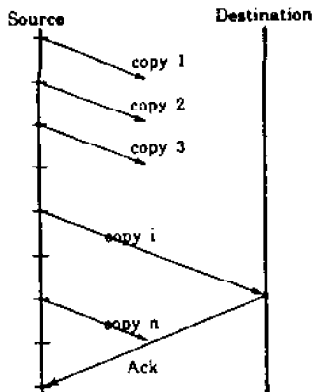

Figure 5: A returning acknowledgment does not tell which copy of the packet is being acknowledged. Using incorrect sample delay can lead either to divergence or to false convergence.

1. $i = n$, Measure sample delay from the last attempt: This removes the positive feedback in the simple timeout algorithm described above, i.e., the measured sample delay does not include the timeout interval. This algorithm converges to wrong values; sometimes causing false alarms. One such example is shown in Figure 6. The actual delay is 15 but somehow the estimate becomes 5. If $k=2$, the timeout occurs at $t_0 = 10$, and the packet is retransmitted. The delay from the last retransmission to the acknowledgment is 5, so the estimate does not change and again the cycle continues.

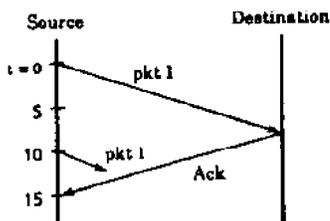

Figure 6: A counterexample showing that measuring round trip delay from the last retransmission may lead to a false convergence. This counterexample also applies if the sample delay is ignored on retransmissions. In both cases, every packet will be retransmitted.

2. $i = 2, 3, \ldots$, Measure sample delay from the second, third, ... attempts: In general, if the acknowledgment is for the $i^{th}$ attempt, measuring it from $j^{th}$ attempt will either lead to divergence (if $i > j$) or false convergence (if $i < j$). The only case for which the procedure converges is if $i = j$. Implementing this solution is rather complex. It would appear on the surface that probably all we need to do is include a copy number in the packet, return it in the acknowledgment, and remember the sending time for each copy. However, even that does not work out. For example, consider a connection with a window size of two. The source sends packets 1 and 2 and remembers the time for sending packet 1 (most networks keep only one timer for a connection). After a while, the sender discovers that packet 1 has timed out. The source resends packet 1 and gets an acknowledgment for the first copy of the packet 2 which also acknowledges receipts of all preceding packets. At this point, it is not possible to say which copy of the first packet is being acknowledged.

Averaging the delay from the first attempt and the delay from the last attempt also leads to similar false convergences or divergence.

At this point, we are tempted to not measure the sample delay for packets with multiple retransmissions. This brings us to our next possibility.

3. Ignore and take no action: Update the delay estimate only if there are retransmissions. This procedure does not have positive feedback. However, like the previous schemes, this one also converges to wrong values sometimes. In fact, the same example as shown in Figure 6 applies. If the delay estimate is 5, when the actual delay is 15, every subsequent packet is retransmitted. The estimate is never updated and remains at 5.

One way to avoid the estimate getting stuck at a low value is to modify this procedure so that the estimate is increased arbitrarily after each case of multiple retransmissions.

4. Ignore the measurement but increase the estimate: There are many ways to increase the estimate. Some of the alternatives that we looked at are:

a. Linear increase:

$$E_i \leftarrow E_{i-1} + \Delta E$$

where $\Delta E$ is a fixed value, for example, 2 seconds.

b. Parabolic increase: The increment is also increased.

$$E_i \leftarrow E_{i-1} + \Delta E_i$$

$$\Delta E_i \leftarrow \Delta E_i + \Delta^2 E$$

Here, $\Delta^2 E$ is a fixed second order increase.

c. Exponential increase:

$$E_i \leftarrow c E_{i-1}$$

Here, $c$ is a dimensionless parameter. It should be greater than one.

d. Exponential increase of the second order:

$$E_i \leftarrow c_i E_{i-1}$$
$$c_i \leftarrow c_i + \Delta c$$

Here, $\Delta c$ is a fixed second order increase in the multiplier.

Of the four alternatives described above, we prefer the third or fourth. The other two use parameters such as $\Delta E$, $\Delta^2 E$ which have dimensions; that is, they are expressed in units of time. A dimensionless parameter is preferable to a dimensional parameter because its range of possible values is much more limited. For example, multiplier $c$ could be set to 2 for the possible range of all delay values. On the other hand, the increment $\Delta E$ in seconds would be inappropriate if the delay itself were in millisecond range. Similarly, $\Delta E$ in milliseconds would be inappropriate if the delay $E$ were in seconds.

All these alternatives also diverge, the maximum allowable loss rate depends on the parameter values.

### Simulation Results

In order to find a suitable solution for the divergence problem, we used a simulation model [9] and compared various alternatives described above. The simulation model allowed us to vary the parameters and traffic conditions. The lessons learned from the simulation are as follows:

1. Let $E_{before}$ and $E_{after}$ be the expected values of the round-trip delay estimate before and after a packet has been retransmitted one or more times. There are three classes of timeout algorithms:

Class I: $E_{after} > E_{before}$

Class II: $E_{after} = E_{before}$

Class III: $E_{after} < E_{before}$

Under sustained loss condition, all class I algorithms diverge because the timeout interval continues to increase. All class III algorithms converge such that the timeout interval reduces progressively towards smaller values and the network gets congested with many duplicate copies of packets. This instability problem due to lower estimate has also been pointed out by Butto, et al [1,2]. With class II algorithms, whenever the estimate becomes sufficiently low it may either converge to zero or oscillate around a low value. In both cases, it results in unnecessary duplicates.

2. All parameters to change the timeout, i.e., the first timeout multiplier $k$, back-off multiplier $b$, and retransmission multiplier $c$, have the same effect, namely increasing them increases the timeout interval. The impact of reducing one can be traded off by increasing another. Back-offs are not necessary, the same effect can be obtained by increasing the timeout multiplier $k$. In fact, one should use $k$ large enough so that second and subsequent retransmissions are rare and back-off parameter is used rarely, if at all.

3. Generally when designing a timeout algorithm or other networking protocols, one should avoid parameters that have dimensions. Parameters without dimension are generally applicable over a wider range of network configurations.

4. The timeout algorithm, the caching algorithm, and window adjustment algorithms used in a network are closely related. A bad timeout algorithm, or a bad caching algorithm may lead to congestion. A timeout is also a indicator of congestion in the network and therefore on a timeout, not only the source should retransmit the packet, but also take action to reduce future input into the network. This leads to a timeout based congestion control policy described in [7].

5. On a timeout, one has a choice of either retransmitting only the packet that timed-out or retransmitting all unacknowledged packets. The latter choice is not a good idea even if the destination is not caching out-of-order packets. This is because timeouts occur when the network is congested and retransmitting too many packets worsens the situation [8].

## Conclusion

The three key ideas presented in this paper are: a layered view of timeout algorithms, the importance of round-trip estimation under sustained loss, and relationship between timeout and congestion control.

Timeout algorithms consists of five procedures which, like layers of a protocol, can be independently chosen and modified. A number of alternatives were presented for each layer.

Of the five layers, the second (estimation under loss) and the fourth (back-off) have not been discussed in literature. We emphasized the importance of the second layer and argued that the fourth layer does not have a significant impact.

The timeouts occur most frequently when the network is congested. Therefore, a good timeout algorithm should perform satisfactorily under congestion and sustained loss. Under low loss conditions, most timeout algorithms perform satisfactorily.

Under sustained loss, all adaptive timeout algorithms discussed in this paper were found to either diverge or converge to values lower than the actual round-trip delay. If an algorithm converges to a low value, it may result in frequent unnecessary retransmissions sometimes leading to network congestion. Divergence is, therefore, preferable in the sense that the retransmissions are delayed.

Timeouts indicate congestion. The input rate to the network should therefore be reduced on a timeout. This leads to a timeout based congestion control scheme, which if used makes back-offs and hence the fourth layer of timeout algorithms unnecessary.

## Acknowledgments

The idea of a different disconnection procedure at setup time is due to Prof. Jerry Saltzer. The back-offs in timeout algorithms, and threshold to bypass gateway failures were first proposed in an algorithm by Geoffrey Cooper.

DECnet is a trademark of Digital Equipment Corporation, and Ethernet is a trademark of Xerox Corporation.